# Conception and FPGA implementation of IEEE 802.11s mesh network MAC layer transmitter

Lamia CHAARI, Rim AYADI and Lotfi KAMOUN

**Abstract**— This paper proposes, a hardware implementation of Wireless Mesh Networks (WMN) medium Access Controller (MAC) layer transmitter. In the literature a lot of works are focused on WMN routing protocol as well as performance analysis and software integration of WMN units, however few works deals with WMN hardware implementation. In this field our contribution is to conceive and to implements on FPGA a WMN MAC transmitter module. Our implementation, written in hardware description language (HDL) is based on the IEEE 802.11 s standard. The hardware implementation retains a good performance in speed.

## 1 INTRODUCTION

THE IEEE 802.11s [1] [2] standard aims to define a MAC and PHY for meshed networks that improve coverage. The mesh is a communication protocol that uses the WiFi existing communication solutions to relay the signal from a device out of covering area toward an access point and then to the IP network. The WMN is a technology for the future networks composed of different types of terminals, including the numeric devices, personal computers and the mobile terminals. Various scenarios can be imagined for the use of WMN networks. They can be used to reach the home networks, to extend the enterprise WLAN networks cover zone and to construct the Ad-Hoc networks. Mesh solutions arrange a very high data transmission rate, allow a fast and simplified deployment, extend the covering area, enhance a strong tolerance to the fails and to interferences and reduce the networks installation costs.

As illustrated in fig.1 [3], the WMN is composed of the following different units:

- Mesh Point (MP): participate in the working of the mesh network and constructs links with neighboring MP.
- Mesh Access Points (MAP): equipped with a function of access point that provides some BSS services to guarantee the communication with the stations in addition to MP functionalities.
- Mesh Portal (MPP): equipped with a bridge function to connect to an external network in addition to MP functionalities.
- Station (STA): outside of the mesh network, connected via MAP.
- Wireless Distribution System (WDS): is used for the data transfer between the MPs, the MPAs and the MPPs.

- *L. Chaari is with Department of Electronic and Information Technology Laboratory (LETI) at Sfax National Engineering School Tunisia.*
- *L. Kamoun is is with Department of Electronic and Information Technology Laboratory (LETI) at Sfax National Engineering School Tunisia.*

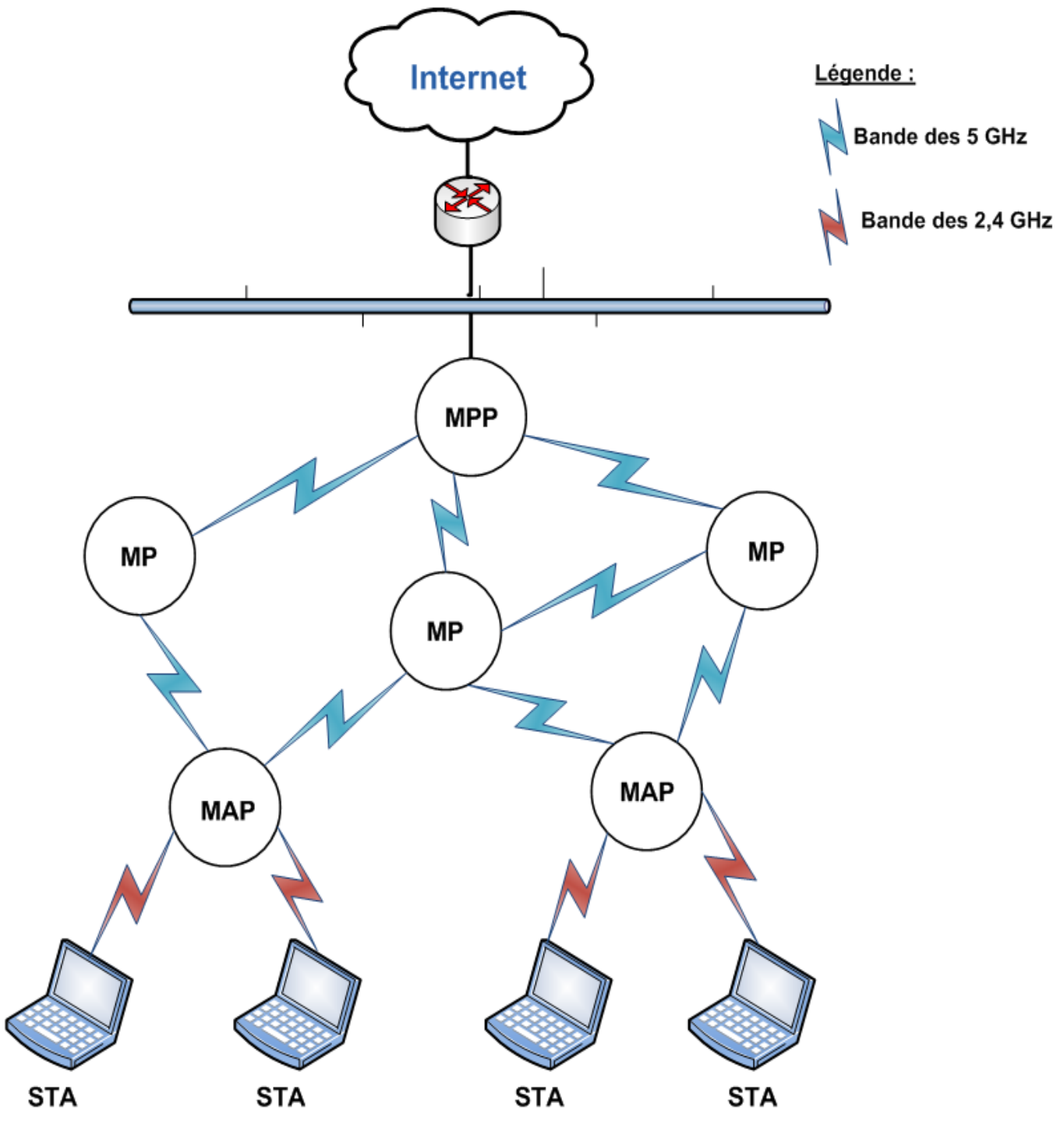

**Fig. 1. WMN Configuration**

The WMN architecture illustrated by the fig.2 [4] .

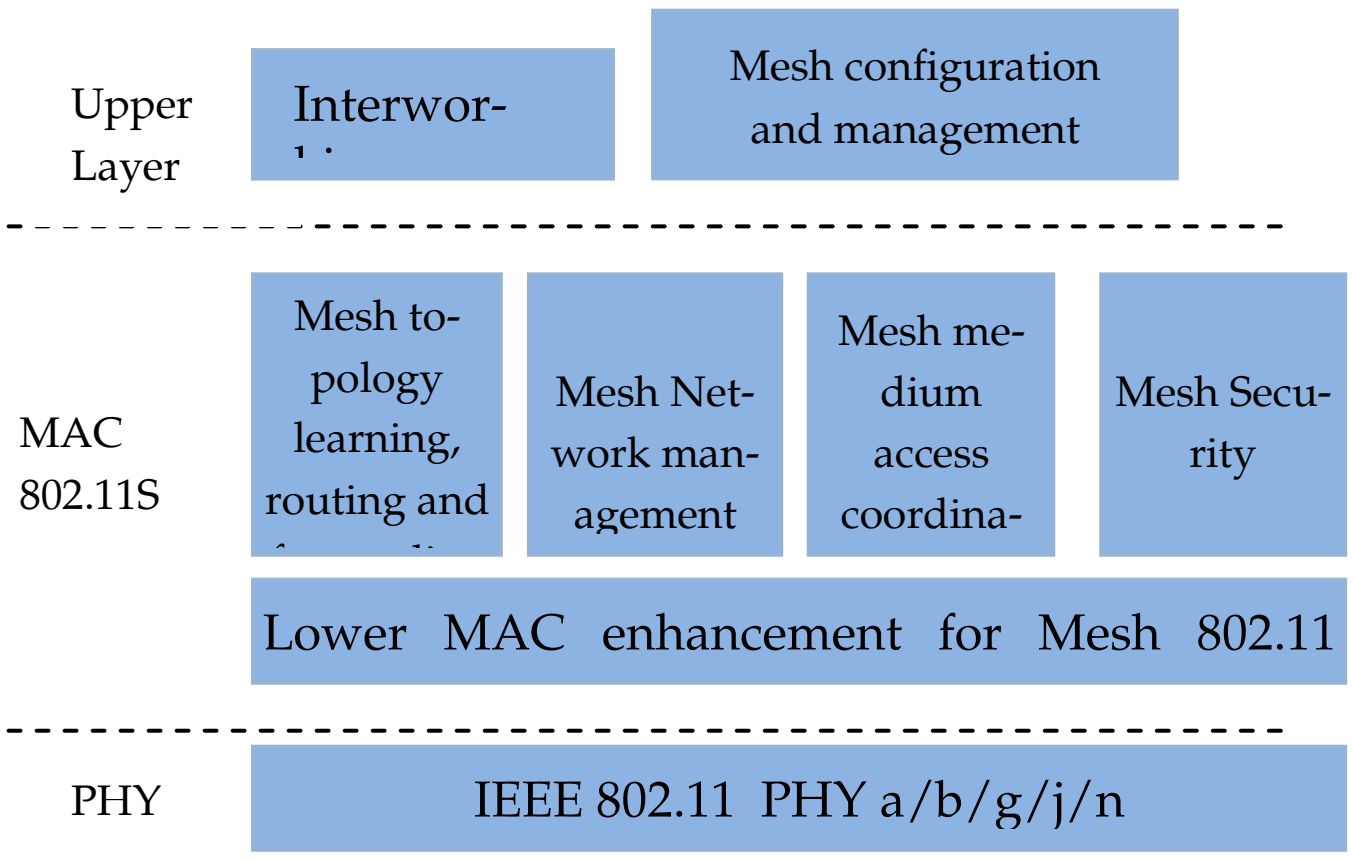

**Fig.2 802.11s basic architecture**

- Mesh topology learning, routing and forwarding block provides discovering neighboring node, getting routing metric and forwarding packet
- Measurement block performs the calculation of routing metric that are used by the routing protocol.
- Medium access coordinator block prevents degraded performance due to hidden and exposed node and performs priority control, admission control and spatial frequency reuse.
- Security block protects data frame. It uses the WLAN security schemes defined in 802.11i.
- Interworking block integrates the wired LAN with the mesh backbone.
- Configuration and management block performs the operation of automatic setting of each MP's Radio Frequency (RF) parameter for QoS management.

After a brief overview of the most important aspects of the developing 802.11s standard, the remainder of this paper is organized as follows. Section II describes frames structure based on the IEEE 802.11s current draft standard D2.0 specification. Section III provides the conception of the WMN MAC transmitter. Section IV validates the functional behavior of each WMN MAC transmitter sub modules and some simulation results are provided. Section V concludes the paper.

## 2 Frames Structure

### 2.1 Mesh Data Frame structure

The 802.11s draft defines new fields in some headers to perform the mesh functionality. A new field called Mesh header, has been introduced in the original 802.11 data field to perform a multi hop routing and is shown in fig.3[1],[5].

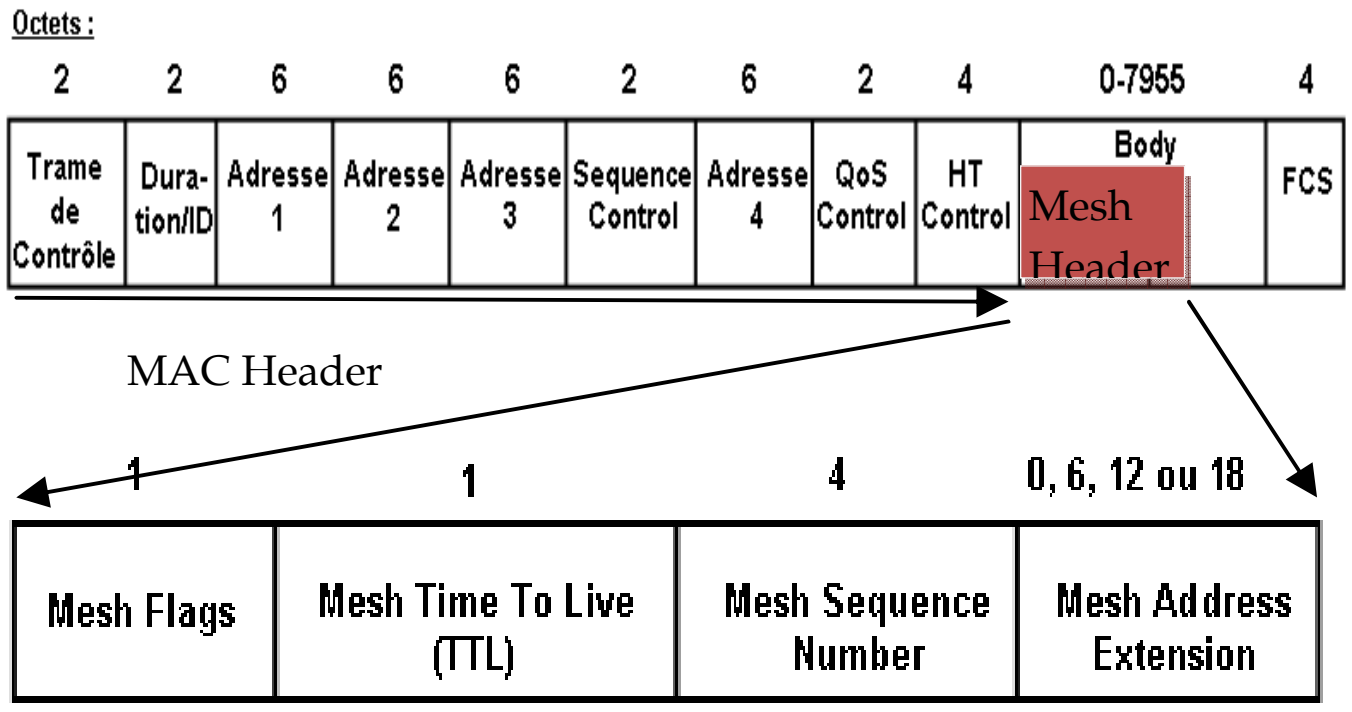

**Fig. 3. Wireless mesh data frame structure**

The mesh header field is a 6 to 24 byte field that includes:
- Mesh Flags field to control mesh header processing,
- Time to live field is used as mechanism to avoid loops in the mesh networks,
- A mesh sequence number to suppress duplicates in broadcast/multicast forwarding and for other services,
- In some cases a 6, 12, or 18-octet mesh address extension field containing extended addresses.

### 2.2 Mesh Management Frames structure

The management frames are used to establish the initial communications between the stations and the access points. So, the management frames provide some services as association, re-association, the authentification, deauthentification and synchronization. The management frame format is illustrated in fig.4.

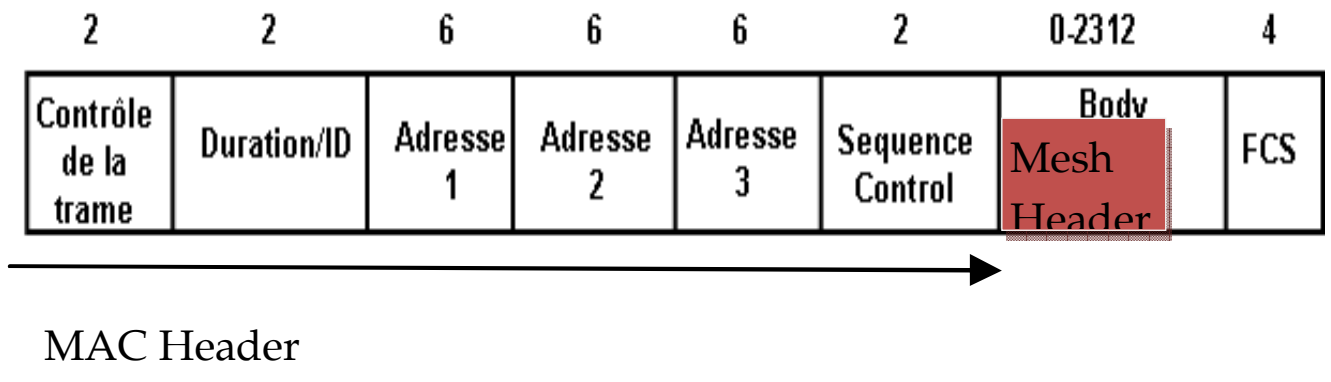

**Fig. 4. Wireless mesh management frame structure**

The address fields for all management frames do not vary by frame subtypes except Multihop Action are as follows.
- The Address 1 (RA=DA) field is the destination of the frame.
- The Address 2 (TA=SA) field is the address of the STA transmitting the frame.
- The Address 3 (BSSID) field of the management frame is determined as follows:
a) If the STA is an AP or is associated with an AP, the BSSID is the address currently in use by the STA contained in the AP.
b) If the STA is a member of an IBSS, the BSSID is the BSSID of the IBSS.
c) In management frames of subtype Probe Request, the BSSID is either a specific BSSID, or the wildcard BSSID.
d) For single-hop management frames transmitted by MPs, the BSSID field is not used and should be set to 0.
The address fields for the management frame subtype Multihop Action are as follows.
The Address 1 (RA) field is the receiver of the frame.

The Address 2 (TA) field is the transmitter of the frame.
The Address 3 (DA) field is the destination of the frame.

There are two new control frames proposed by the IEEE request to switch (RTX) frame and the clear to switch (CTX) frame. These frames are used to perform backhaul channel change operations.

## 3 WMN MAC TRANSMITTER CONCEPTION

The conceived WMN MAC Transmitter is composed with four main sub-modules: transmission control, allocation control, frame transmission and frame computing.

-Transmission control sub-module permits to control the different data transmission steps. It arbitrates the other sub-modules of the MAC 802.11s transmitter. It is activated by five signals MSDURDY, REC_DATA, REC_CTS, REC_RTS and REC_ACK.

- Frame computing sub-module: permits to build the frame based on the received frame subtype information. It is activated by the EN_BUILDFRAME signal which is received from the transmission control sub-module.

- Frame transmission sub-module is activated by the signal FRAGMENT. It transmits the different frame fields the TX_LINE output. Finally a TRANSMIT_COMPLETE signal is sent to the Transmission control sub-module.

- Allocation control sub-module is activated by the ENABLE_MEDIUM and ENABLE_RETRY signals. When it is activated by the ENABLE_MEDIUM signal, it verifies the availability of the medium by the verification of the CARRIER_SENSE signal of the physical layer and the value of NAV that is provided by the NAV register. If the medium is available it sends the ACCESS_GRANTED signal to the control transmission sub-module. If the medium is not available it waits for a certain time that is determined by the BACKOFF ALGORITHM.

When this sub-module is activated by ENABLE_RETRY, first it increments the tentative counter and then it verifies the availability of the medium and the process continues as describes above.

The sequence diagram presented by fig.5, illustrates the different signals exchanged between WMN MAC Transmitter sub-modules.

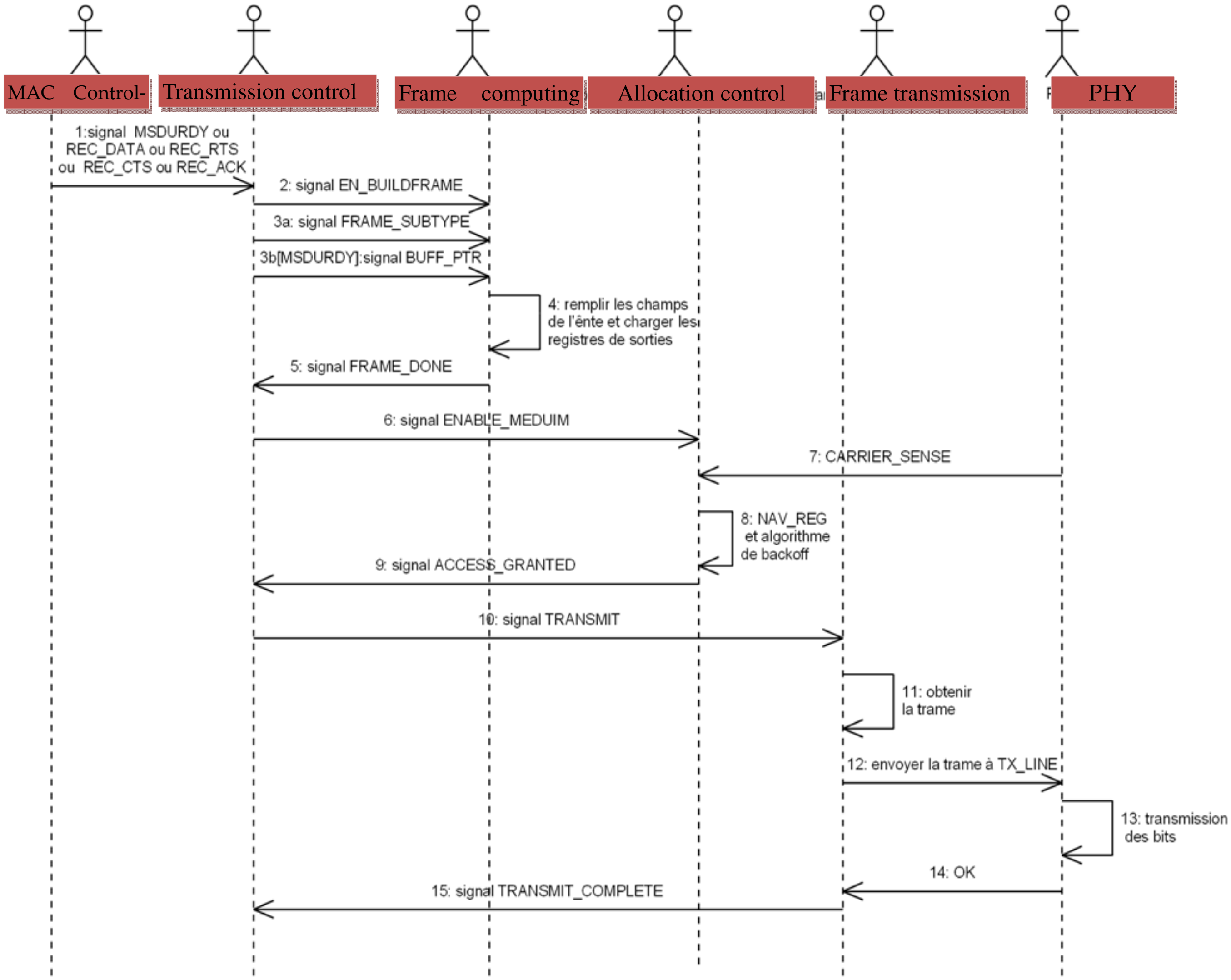

**Fig. 5. Sequence Diagram of a MAC 802.11 transmitter**

## 4 WMN MAC TRANSMITTER FPGA IMPLEMENTATION

In order to validate the functionalities of the conceived transmitter, we have used « ModelSim XE-III » which includes a complete HDL simulation environment that allows us to verify the functional behaviour, the VHDL code source and the synchronization models of our conceptions. Also it is optimized for the use with all Xilinx ISE™ products.

### 4.1 Transmission control sub-module

If the signal rec_data is high the signal en_buildframe will be set to 1 and the signal frame_subtype will be equal to 101011 to indicate that the frame that is going to be constructed is ACK type. When it receives the signal frame_done in the high state, it sends the signal en_meduim. In the same way, when it receives the signal access_granted in the high state, it sends the signal transmitted and it remains waiting to the signal transmit_complete in the high state. After, all these signals are going to be reset. The following fig.6 illustrates the functional behavior of this block in this case.

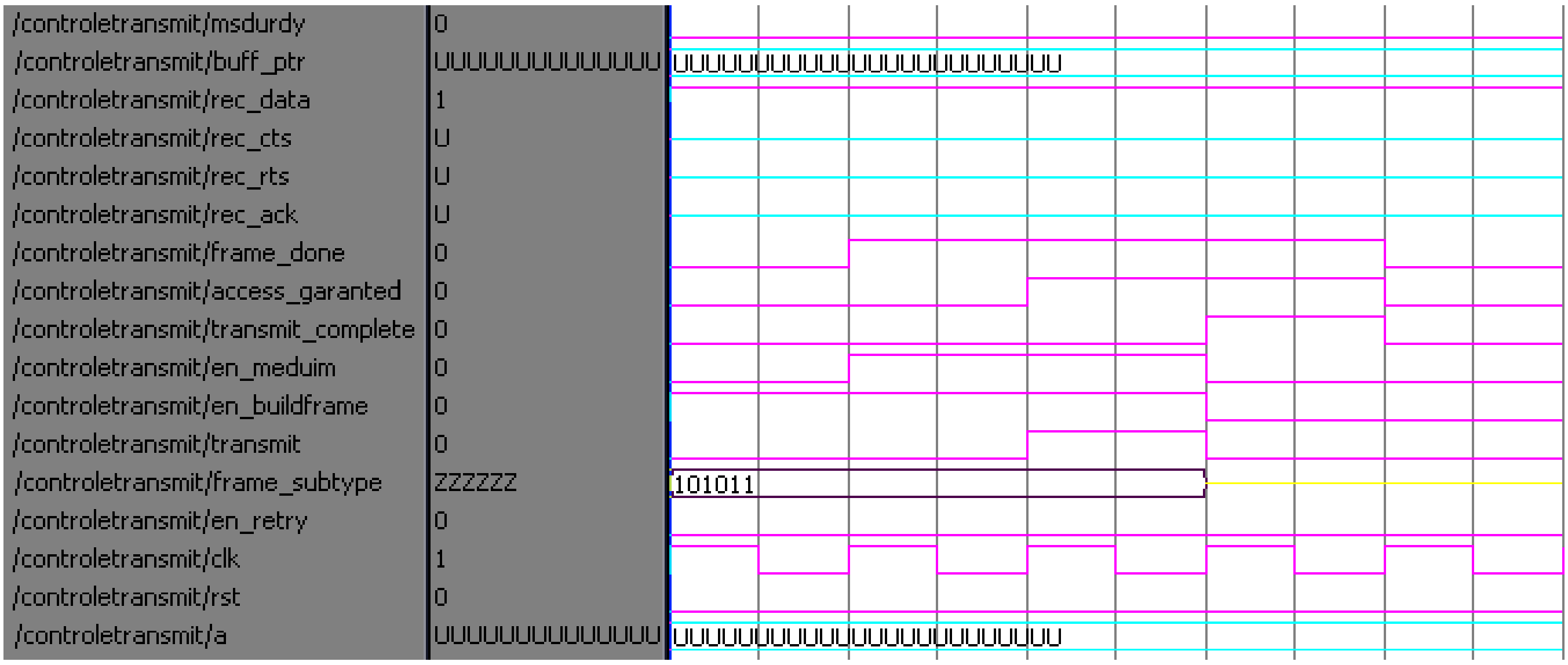

**Fig. 6. Transmission control sub-module functional behavior when (rec_data = '1')**

If the signal rec_cts is high the signal en_buildframe will be set to 1 and the signal frame_subtype will be equal to 010000 to indicate that the frame that is going to be constructed is data frame. When it receives the signal frame_done in the high state, it sends the signal en_meduim. In the same way, when it receives the signal access_granted in the high state, it sends the signal transmitted and it remains also in waiting to the signal transmit_complete in the high state. After, all these signals are going to be reset and the block remained in waiting of the signal rec_ack. The following fig.7 illustrates the functional behavior of this block in this case. In the case where this signal is at the low state the value of the signal in _retry is going to be equal to '1'.

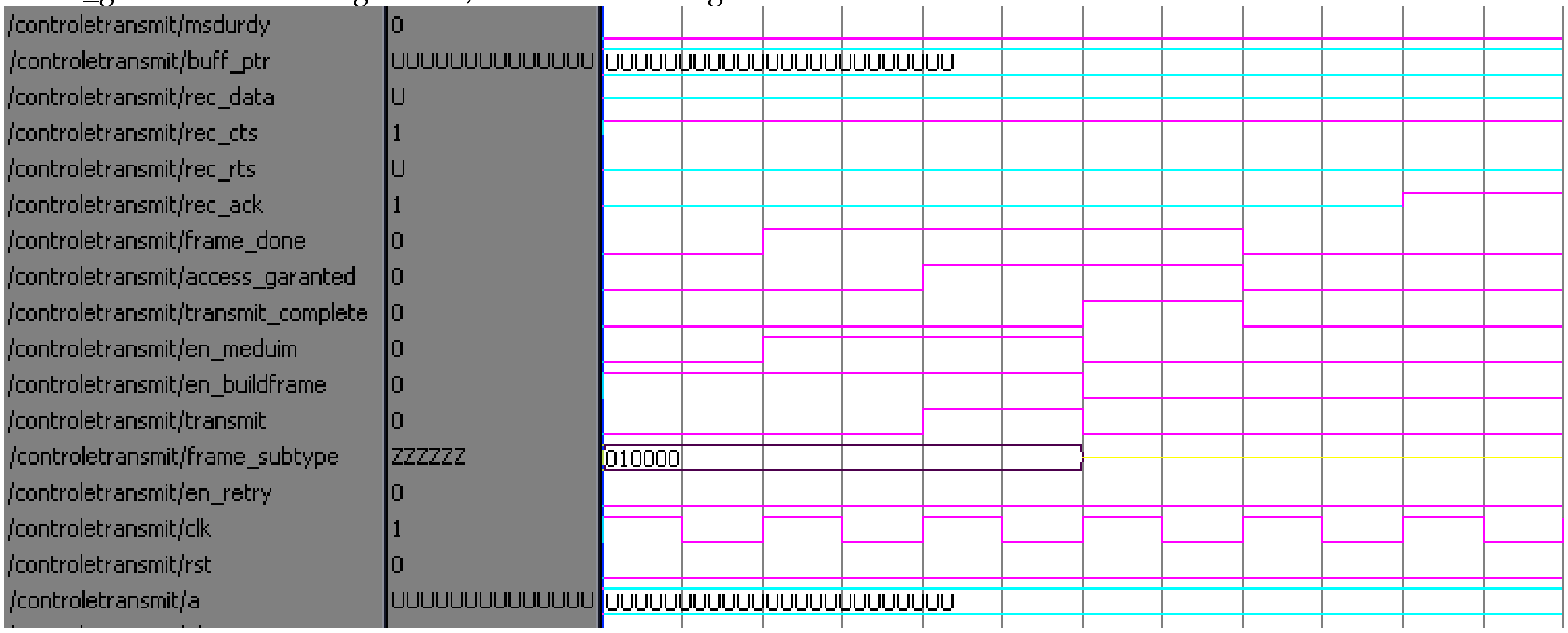

**Fig. 7. Transmission control sub-module functional behavior when (rec_cts = '1')**

If the signal rec_rts is high the signal en_buildframe will be set to 1 and the signal frame_subtype will be equal to 010011 to indicate that the frame that is going to be constructed is CTS frame. When it receives the signal frame_done in the high state, it sends the signal en_meduim. In the same way, when it receives the signal access_granted in the high state, it sends the signal transmitted and it remains also in waiting to the signal transmit_complete in the high state. After, all these signals are going to be reset and the block remained in waiting of the signal rec_data. The following fig.8 illustrates the functional behavior of this block in this case.

If the signal msdurdy is high the signal en_buildframe will be set to 1 and the signal frame_subtype will be equal to 101101 to indicate that the frame that is going to be constructed is RTS frame. When it receives the signal frame_done in the high state, it sends the signal en_meduim. In the same way, when it receives the signal access_granted in the high state, it sends the signal transmitted and it remains also in waiting to the signal transmit_complete in the high state. After, all these signals are going to be reset and the block remained in waiting of the signal rec_cts. The following fig.9 illustrates the functional behavior of this block in this case.

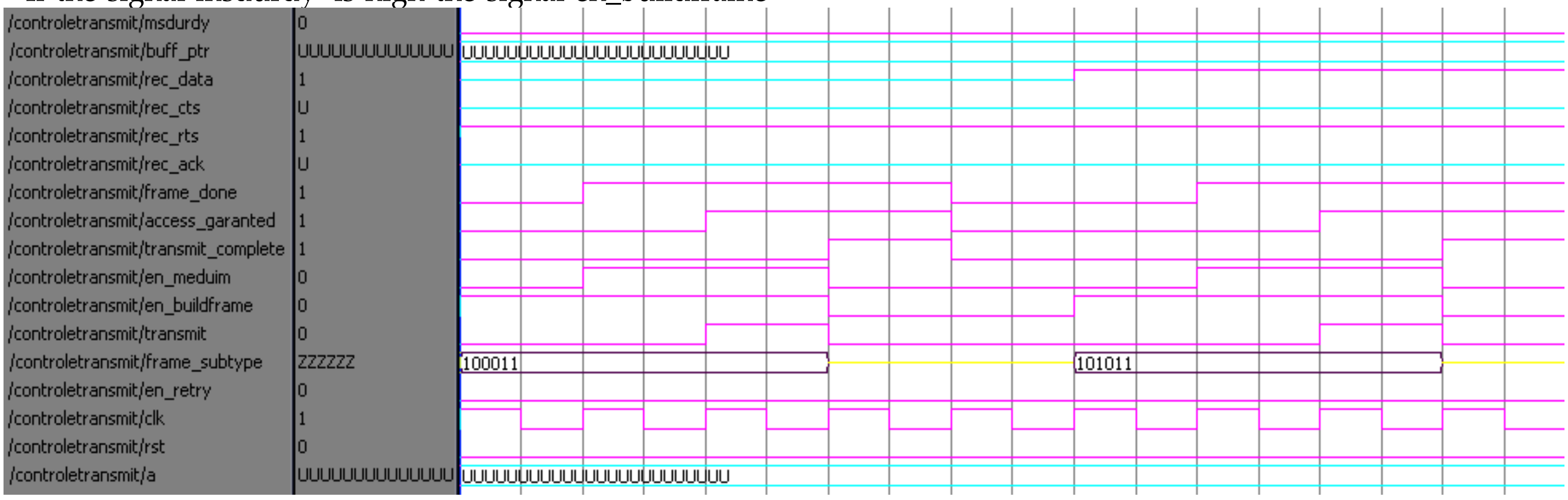

Fig. 8. Transmission control sub-module functional behavior when (rec_rts = '1')

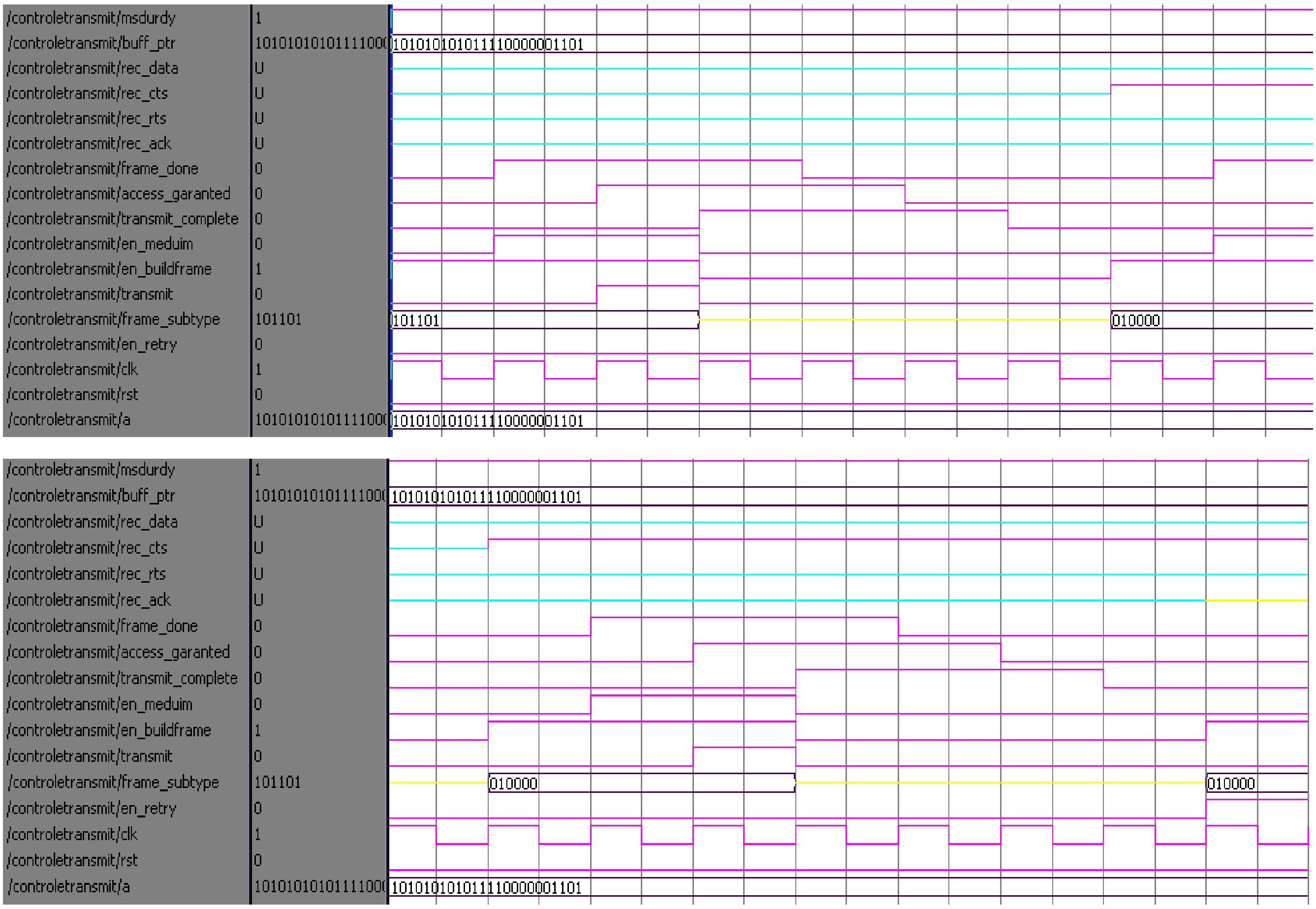

Fig. 9. Transmission control sub-module functional behavior when (msdurdy = '1')

### 4.2 Frame computing sub-module

The frame computing sub-module is enabled by the EN_BUILDFRAME signal from the Transmission control sub-module. Based on the FRAMESUBTYPE input, the frame is identified. The most frame computing sub-module entities are: Control Header, DID, Generate Address and Frame Check sequence which builds different parts of frames header.

- The Frame Control Header entity sets the following values (Protocol version (2) 00, Type (2) Input, Subtype (4) Input ToDS (1), FromDS (1), MoreFragements (1) ,Retry (1) Input Power Management system (1), MoreData (1),WEP (1), Order (1). These values are stored in the register and then sent to the Frame Transmission sub-module through FCH signal. Fig.10 illustrates the functional behavior of this entity.

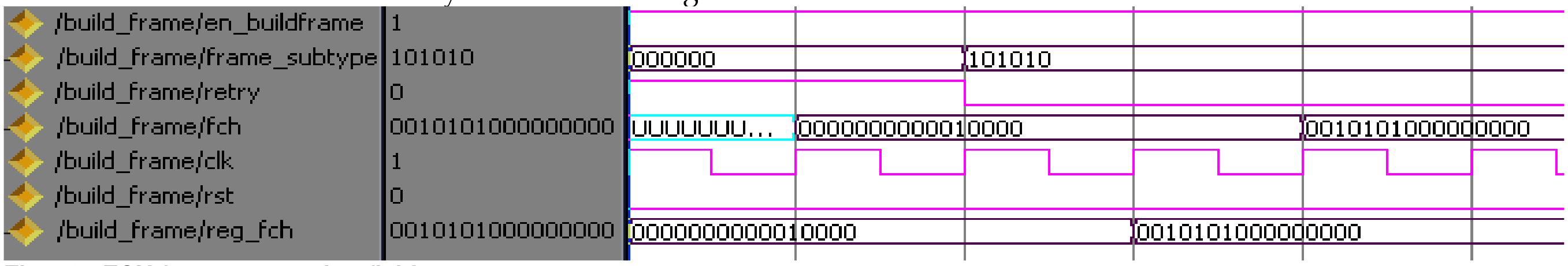

**Fig. 10. FCH frame computing field**

The duration/ID entity takes the input from the NAV register and sets the DID value, then the values are sent to the frame transmission sub-module. If the frame type is PS_POLL (frame_subtype=100101) then the value of the signal reg_did is going to be the value of nav_reg while replacing the last two bits by '1'. If the frame type is CFP type (frame_subtype=010110) then the value of reg_did will be equal to "0000000000000001" and the DID signal receives it otherwise the signal reg_did is going to be the nav_reg value while replacing the last bit by '0'. Fig.11 illustrates the functional behavior of this entity.

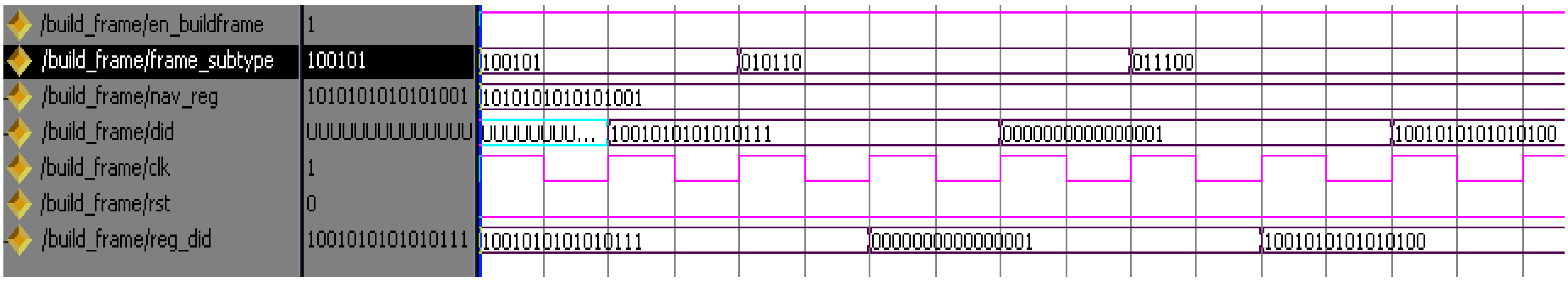

**Fig. 11. Duration/ID frame computing field**

- The Address Generation entity based on frames subtype, it takes BUFF_PTR input from the buffer and stores the generated addresses in a register. The generated addresses are sent to the Frame Transmission sub-module block through the four signals ADDR1, ADDR2, ADDR3 and ADDR4.

- Frame Sequence Control entity generates frame sequence only for data frames. It checks whether the frame has to be fragmented or not. If the frame has to be fragmented this entity sends a FRAGMENT signal to the Frame Transmission sub-module. It increments the sequence counter for each successive frame. The values are stored in the register and then sent to the Frame Transmission sub-module through FCS signal. Fig.12 illustrates the functional behavior of this entity.

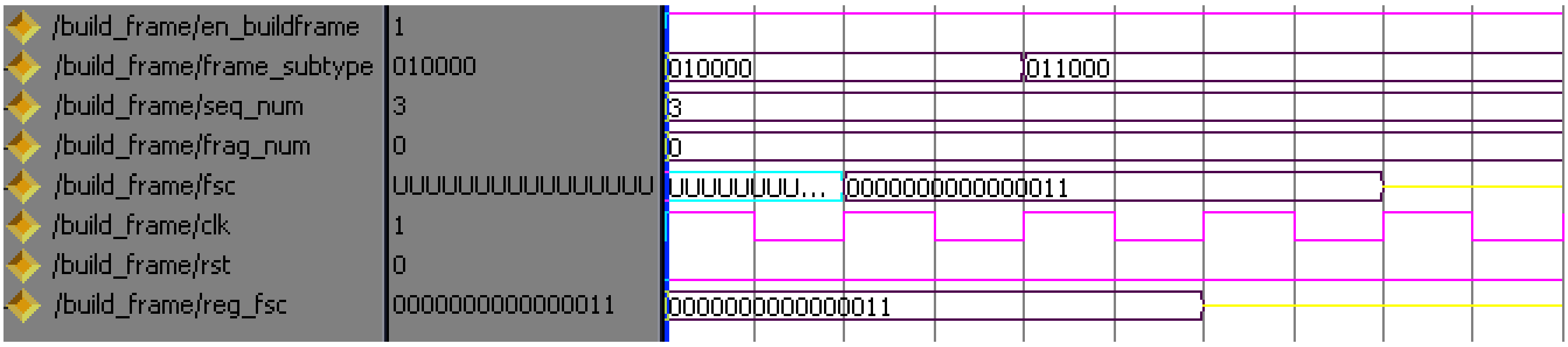

**Fig. 12. FSC computing field**

### 4.3 Transmission sub-module

It has a multiplexer and a 32-bit shift register. FCH, DID, ADDR1, ADDR2, ADDR3, ADDR4, FCS, DATA are all given as inputs to the multiplexer. It also has a 4-bit select line. The output of the multiplexer is sent to the 32-bit shift register.

The different parts of the frame are transmitted through TX_LINE signal. Fig.13 demonstrates the functional behavior of this entity.

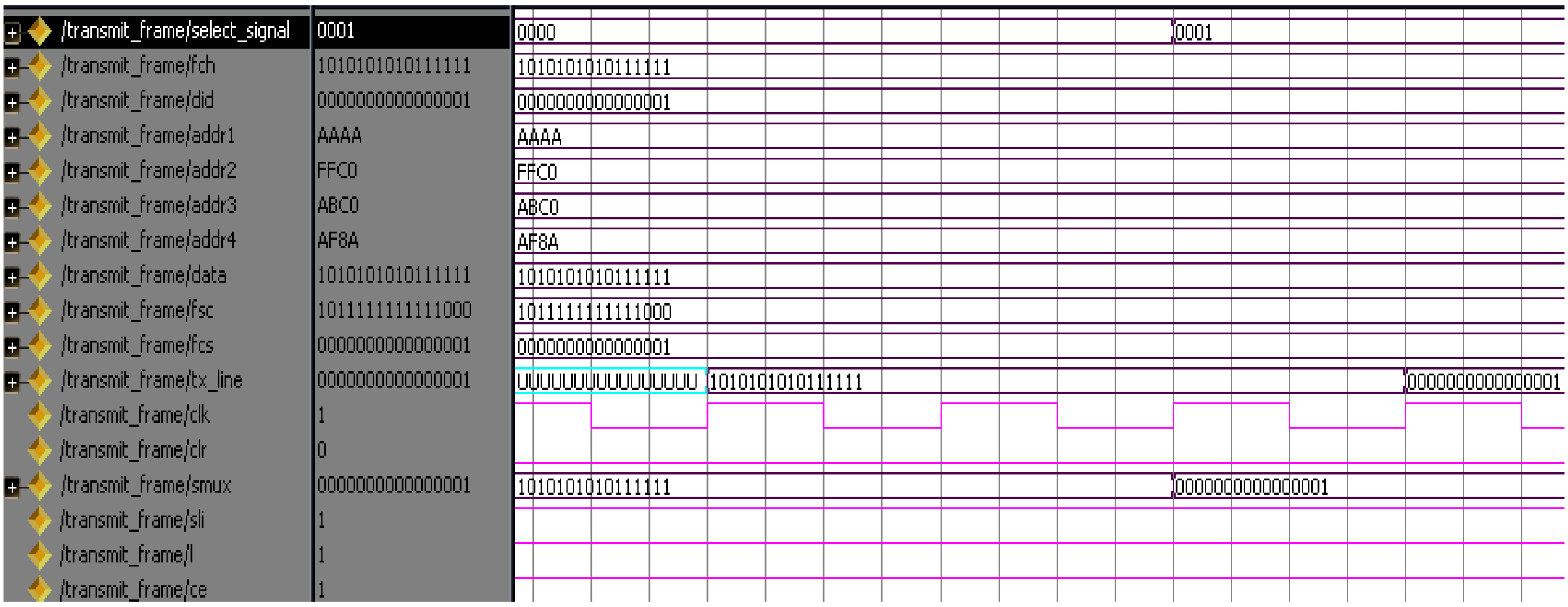

**Fig. 13. Frame transmission sub-module functional behavior**

### 4.4 Allocation control sub-module

The allocation control sub-module contains three blocks: " collisions avoidance entity ", "retry counter entity "and" BACKOFF entity ". The "retry counter entity" maintains the attempts number. When it is activated by the ENABLE_MEDIUM signal, the "collisions avoidance entity" verifies the medium availability by CARRIER_SENSE signal verification (from the physical layer) and the value of NAV that is provided by the NAV register. If the medium is available it sends the ACCESS_GRANTED signal to the «transmission controls sub-module ". If the medium is not available it waits for a certain time that is determined by the " BACKOFF entity ". It verifies again the availability of the medium after the tentative counter. Fig illustrates the

#### *4.4.1 Backoff entity*

The Backoff entity is used to calculate a backoff random that is decreased when the medium is inactive. This entity is activated by avoidance collision entity when the medium is occupied. This entity is activated by the EN_BACKOFF signal. The operation of this entity starts with CW(Contention Windows) calculation using SSRC and SLRC values that are defined in every station. These two values are attached to the emission attempts counters SRC (Shorts Retry Count) and LRC (Long Retry Count). These counters are initialized in the following cases: CTS reception after RTS emission, ACK reception after data frame emission and "Broadcast" or" Multicast " frame reception. And if they reach their limit then the frame will be rejected. The value of CW is provided to " Random Number Generator" that produces a random number in the [0, CW] interval. Then the time of backoff will be calculated like follows:

Backoff Time = Random () * SlotTime with Random () is a value in the interval [0, CW] where CW to take a value between CWmin and CWmax. These last two values are defined according to the physical layer. In conclusion, the time of backoff is calculated and it is sent on BACKOFF_VAL output. Fig.14 demonstrates the functional behavior of this entity.

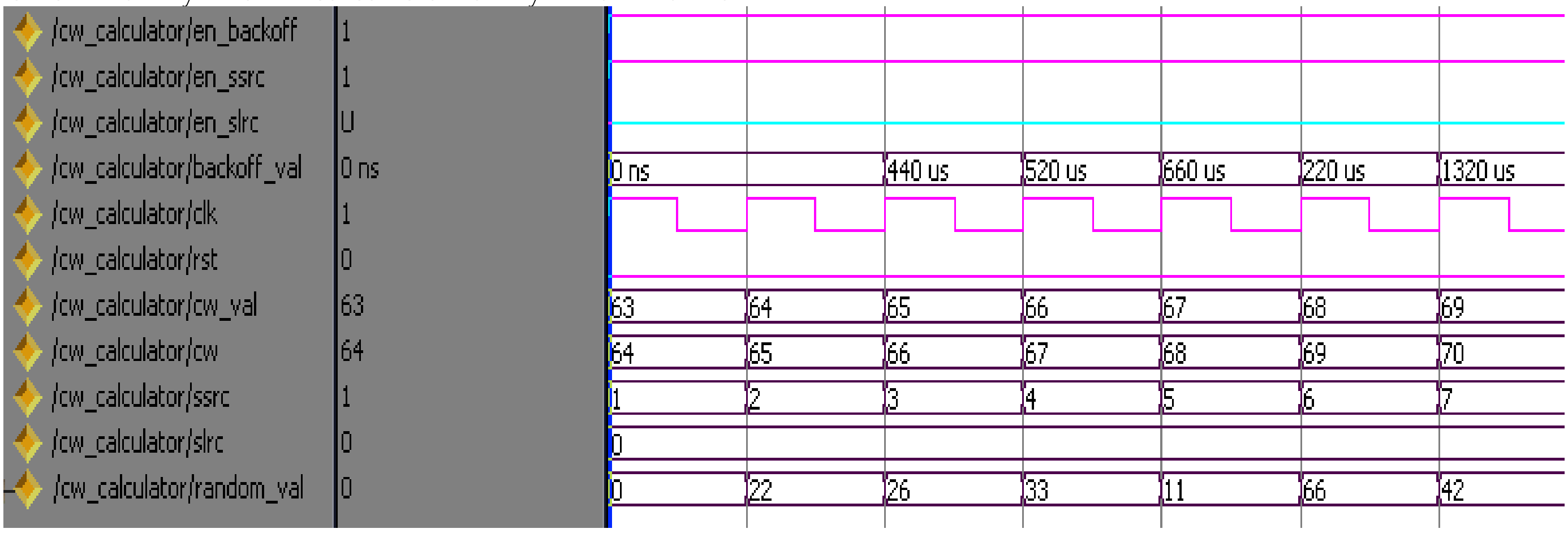

**Fig. 14. Backoff entity functional behavior**

#### *4.4.2 Counter entity*

This entity is a counter that is activated by start_count signal. The counter is initialized when reset signal is set to high level. Fig.15 demonstrates the functional behavior of this entity.

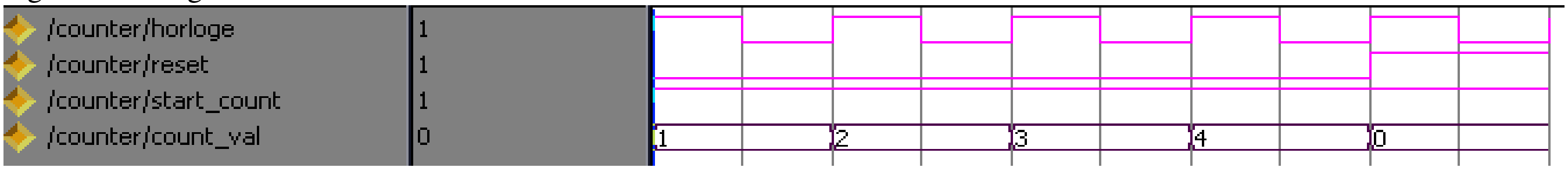

**Fig. 15. Retry counter entity functional behavior**

### *4.4.3 Collisions avoidance entity*

This entity is activated by en _medium or en_retry the signasl and its output is the access_granted signal by which it indicates if the medium is available (access_granted <= '1') or no (access_granted <= '0'). To activate backoff entity, this block sets en_backoff in the high state and to activate retry counter entity, it also sends sets the start_count in the high state. If the nav-reg value is equal to zero and if the carrier_sense <= '0' then the medium is available to accept frames. If the nav_reg value is different from 0, the access to the medium is forbidden. Then the attempts number is incremented and backoff algorithm is executed. If the nav_reg value is equal to zero but the value of carrier_sense is equal to 1, then the signal access_granted will be equal to zero this is illustrated in the fig.16a. In this case, retry counter entity is activated in order to increment the attempts number. If this number passes the limit dot11rst_threshold (for example it is equal to 10 in our work) then access_granted is going to be equal to 0, this is illustrated in the fig.16b. Otherwise the backoff entity is activated and the signal access_granted is equal to 1 only after the backoff period which is determined by the backoff_val signal, this is illustrated in the fig.16

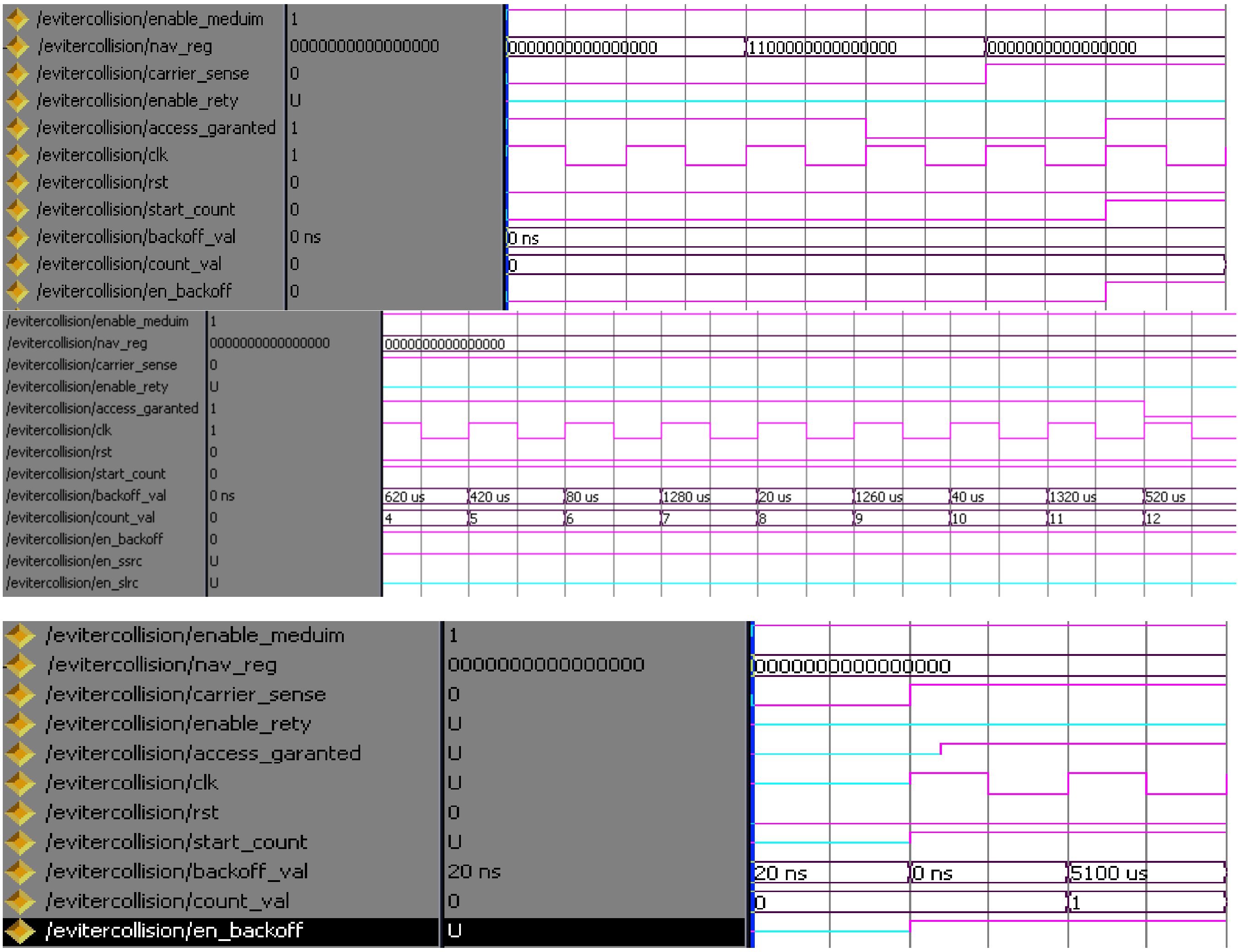

**Fig 16 Collision avoidance functional behavior**

## 5 Conclusion

This work focuses on the problem of simultaneously designing and implementing Wireless Mesh Networks (WMN) medium Access Controller (MAC) layer transmitter. The proposed architecture is modular. Based on the simulation the design was found to conform to the design specifications.

As future work we can:

- improve our implementation by adding advanced functionalities such as Mesh topology learning, routing and forwarding, mesh security and mesh network management.
- implement the (WMN) medium Access Controller (MAC) layer receiver

**Lamia CHAARI** was born in Sfax, Tunisia, in 1972. She received the engineering and PhD degrees in electrical and electronic engineering from Sfax national engineering school (ENIS) in TUNISIA. Actually she is an assistant professor in multimedia and informatics higher institute in SFAX She is also a researcher in electronic and technology information laboratory (LETI). Her scope of research are communications, networking and signal processing which are specially related to wireless and new generation networks.

**Lotfi Kamoun** was born in Sfax Tunisia, 25 January. 1957. He received the electrical engineering degree from the Sciences and Techniques Faculty in Tunisia. Actually he is a Professor in Sfax national engineering school (ENIS) in TUNISIA. He is the director of electronic and technology information laboratory (LETI). His scope of research are communications, networking, Software radio and signal processing which are specially related to wireless and new generation networks.